# Rovibrational Polaritons in Gas-Phase Methane

Adam D. Wright, Jane C. Nelson, Marissa L. Weichman*

Department of Chemistry, Princeton University, Princeton, New Jersey 08544, United States

**ABSTRACT:** Polaritonic states arise when a bright optical transition of a molecular ensemble is resonantly matched to an optical cavity mode frequency. Here, we lay the groundwork to study the behavior of polaritons in clean, isolated systems by establishing a new platform for vibrational strong coupling in gas-phase molecules. We access the strong coupling regime in an intracavity cryogenic buffer gas cell optimized for the preparation of simultaneously cold and dense ensembles and report a proof-of-principle demonstration in gas-phase methane. We strongly cavity-couple individual rovibrational transitions and probe a range of coupling strengths and detunings. We reproduce our findings with classical cavity transmission simulations in the presence of strong intracavity absorbers. This infrastructure will provide a new testbed for benchmark studies of cavity-altered chemistry.

## 1. INTRODUCTION

Optical cavities facilitate resonant interactions between a photonic mode and a transition of intracavity matter, such as atoms, molecules, quantum dots or semiconductor material.[1-6] Within the strong coupling regime, the rate of coherent excitation exchange between the optical cavity mode and intracavity material overwhelms that of all dissipative processes (Figure 1a).[2,7-9] This regime manifests in the appearance of a doublet structure in the cavity transmission spectrum[10,11] whose energetic separation, known as the vacuum Rabi splitting, $\Omega_R$, is proportional to the light-matter coupling strength and can arise without induction by an external field.[10] These new hybrid light-matter eigenstates are termed polaritons, and are analogous to Autler-Townes dressed states, albeit driven by fluctuations in the cavity vacuum field rather than an external oscillating field.[12,13] Here, we report on a new platform to achieve vibrational strong coupling in cold, gas-phase molecules, and demonstrate strong coupling of a single rovibrational transition of methane. We envision that this gas-phase infrastructure will ultimately provide a testbed to interrogate chemical processes under cavity strong-coupling with quantum-state specificity and without the complications of solvent.

The interplay between the local and collective dynamics of an ensemble of molecules in dialog with a common field of light is a rich sandbox for chemistry. It has become clear in recent years that cavity-coupled molecules may feature energetics, reactivity, and photochemistry distinct from their free space counterparts.[1,3,6,14-20] Strong coupling of molecular vibrations, in particular, offers a tantalizing approach to achieve mode-selective chemistry without direct laser excitation.[3,16] Vibrational strong coupling (VSC) was first demonstrated in Fabry-Pérot cavities in 2015 by the groups of Ebbesen[21] and Simpkins.[22,23] The influence of VSC on chemistry has subsequently become a topic of considerable interest. There is a growing body of experimental work showing reduced reaction rates,[24-26] and modified branching ratios[27] and equilibrium constants[28] of thermal, solution-phase reactions under VSC. While these observations of cavity-altered chemistry are compelling, their mechanisms and reproducibility remain a subject of intense discussion in the community.[16-19,29,30] Experimental VSC work has thus far tackled condensed-phase systems where solvent effects are not well understood, mechanisms are not easily rationalized, and detailed theoretical treatment remains challenging.[16,17,31,32] Convergence between theoretical and experimental efforts in polariton chemistry requires laboratory work on simple, clean systems. Moreover, the polariton chemistry community has repeatedly cited the need for fundamental tests of polariton behavior in the gas phase.[5,18,33,34]

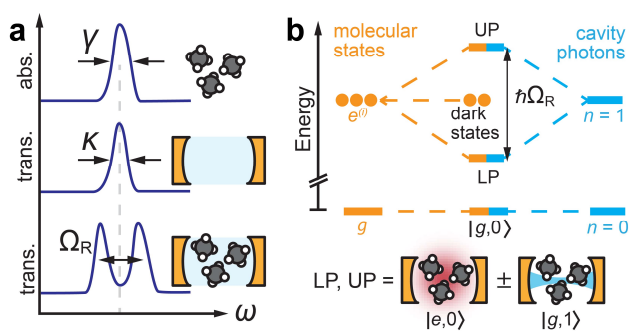

**Figure 1**. (a) Schematic of strong coupling between a molecular transition with linewidth $\gamma$ and a cavity mode with photon loss rate $\kappa$. When the rate of resonant exchange of photons, $\Omega_R$, exceeds $\gamma$ and $\kappa$, two new resonances can be resolved in the cavity transmission spectrum. (b) Energetic diagram of $N$ molecules coupled to a cavity mode within the Tavis-Cummings model. The molecular ensemble can either be found in the collective ground state, $g$, or in one of $N$ excited states where the $i$th molecule is excited, $e^{(i)}$. In this example, the cavity is either dark ($n = 0$) or populated with one photon ($n = 1$). The collective molecular excited state hybridizes with the $n = 1$ photonic state to yield lower and upper polaritons (LP, UP). $N-1$ dark states remain at the original molecular energy.

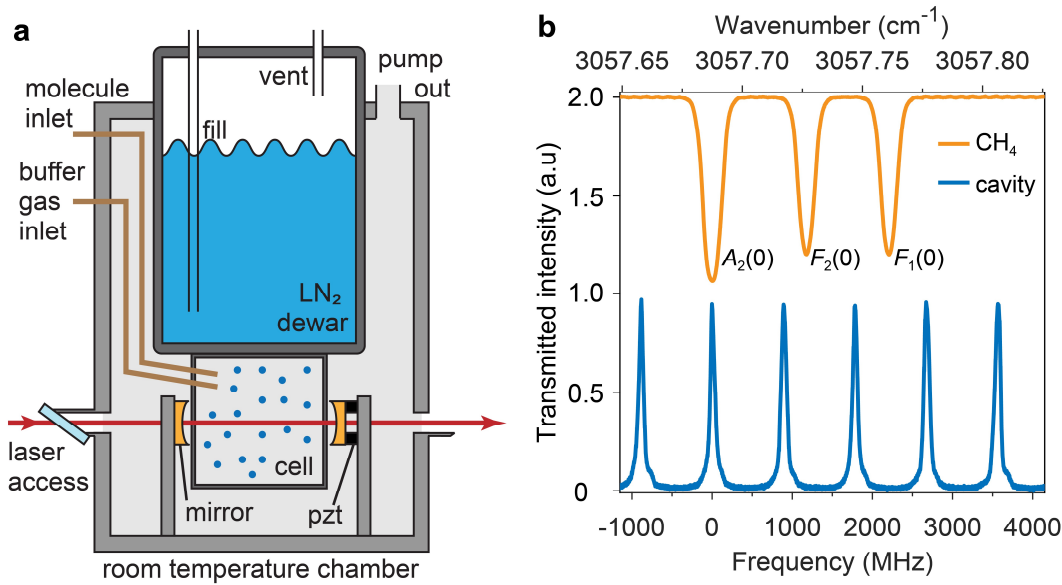

**Figure 2.** (a) Cryogenic buffer gas cell used to prepare a cold, dense molecular sample. A Fabry-Pérot optical cavity encloses the cell to achieve in situ strong coupling. (b) Experimental transmission spectrum of an empty near-confocal cavity (blue) and transmission spectrum of $CH_4$ at 120 K with a number density of $1.5 \times 10^{15}$ cm$^{-3}$ (orange, vertically offset by 1 a.u. for clarity). The frequency axis is referenced with the $\nu_3$ $J = 3 \to 4$ $A_2(0)$ resonance of $CH_4$ located at 3057.687423 cm$^{-1}$ corresponding to 0 MHz. The cavity transmission spectrum is normalized against the amplitude of a fringe far from resonance.

Here, we take a first step to bridge this gap by developing a gas-phase platform for vibrational molecular polaritons. Although strong coupling of electronic transitions has been observed in cold atoms,[7,35-37] polaritons have not, to our knowledge, been previously reported in gas-phase molecules. Low number densities and small oscillator strengths have hindered the demonstration of strong light-matter coupling of gas-phase molecular vibrations, since the collective Rabi splitting scales linearly with the transition dipole moment and with the square root of the concentration of cavity-coupled molecules, according to the Tavis-Cummings model (Figure 1b).[1,4,5]

By definition, reaching the strong coupling regime requires only that $\Omega_R$ exceed the molecular absorption and optical cavity mode linewidths. To maximize $\Omega_R$ and minimize the linewidths for our gas-phase molecular ensemble, our strategy is to prepare a sample that is both cold and dense to (a) narrow the molecular Doppler linewidth and (b) reduce the molecular partition function to maximize population in low-lying rovibrational states, while simultaneously (c) maximizing the intracavity molecular number density. Here, we target cavity-coupling of methane ($CH_4$) as a stable, closed-shell gas-phase species featuring well-characterized transitions and a small rovibrational partition function. We show that when the cavity is coupled to a transition arising out of a highly-populated low-lying rovibrational state, the collective Rabi splitting can exceed the Doppler linewidth even with relatively modest number density.

## 2. EXPERIMENTAL SECTION

Our experiment rests on the use of a cryogenic buffer gas cell (CBGC) for preparation of the gas sample, combined with a feedback-stabilized Fabry-Pérot optical cavity and probed using continuous-wave mid-infrared spectroscopy. Our intracavity CBGC draws inspiration from similar implementations used for cavity-enhanced spectroscopy of cold molecules.[38,39] We describe this apparatus here briefly; additional experimental details are provided in Section S1 of the Supporting Information (SI).

Our homebuilt CBGC and vacuum chamber setup is shown in Figure 2a, with a 3D model depicted in Figure S1. We rely on collisions with a cold inert buffer gas inside the CBGC as a near-universal method for molecular cooling.[40] Warm methane molecules flow into a cell anchored to a liquid nitrogen dewar, where they are thermalized to the cell wall temperature via collisions with helium buffer gas. For all experiments reported here, we use a 100 sccm flow rate of helium buffer gas, which corresponds to a 260 mTorr background pressure. All our measurements are consistent with molecular thermalization to 120 K.

An in-vacuum Fabry-Pérot cavity is constructed around the CBGC. This cell geometry is better suited for strong coupling for this initial demonstration than, for instance, a molecular beam, as we can work with higher molecular number densities and more flexible cavity lengths. Here, we use a near-confocal cavity geometry with length $L \approx 8.36$ cm matched to the mirror radii of curvature. Confocal cavities feature degenerate transverse spatial modes, allowing all spatial modes to be brought simultaneously in resonance with the target molecular transition.[36,37] The cavity consists of two plano-concave mirrors, fabricated by depositing a nominal thickness of 11 nm of gold on 1" diameter $CaF_2$ substrates to achieve a mirror reflectivity of 87.8% at 3270.4 nm (3057.7 cm$^{-1}$), yielding a cavity finesse of ~24. We plot the transmission spectrum of an empty near-confocal cavity in Figure 2b, which features a cavity linewidth of 65 MHz full width at half-maximum (fwhm) and a mode spacing of 895 MHz. The cavity length is actuated using a ring-shaped piezo-electric chip. We actively stabilize the cavity length via a side-of-line lock to a cavity fringe at a wavelength near 1550 nm, using a stable metrology-grade continuous wave laser operating near 1550 nm as a frequency reference.

We probe the cavity transmission using a mid-infrared continuous-wave distributed feedback interband cascade laser with a central frequency tunable over 3262 to 3278 nm (3065 to 3050 cm$^{-1}$). We mode-match the beam to the TEM$_{00}$ cavity mode and detect the intensity of light transmitted through the cavity while sweeping the laser frequency. Our full laser table layout is shown in Figure S2.

## 3. RESULTS AND DISCUSSION

Here, we focus on achieving cavity-coupling of the $J = 3 \rightarrow 4$ multiplet of methane's $\nu_3$ C–H stretching fundamental lying near 3.27 μm (3058 cm$^{-1}$).[41] The $\nu_3$ mode is triply degenerate, implying that its excitations can couple with the cavity field regardless of molecular orientation. However, a CH$_4$ molecule must reside in the $J = 3$ rotational level of the vibrational ground state in order to undergo resonant cavity-coupling in our system. The absorption spectrum of CH$_4$ in this region is shown in the upper trace of Figure 2b, as measured for a sample cooled in the CBGC with cavity mirrors removed. The CH$_4$ transitions feature Gaussian Doppler-broadening to 180 MHz fwhm, commensurate with a translational temperature of 120 K. Pressure and transit-time broadening are comparably negligible (1.6 MHz and 0.5 MHz, respectively) under our conditions.

To engineer strong light-matter interactions in methane, we lock the cavity length to keep one fringe resonant with the target $\nu_3$ $J = 3 \rightarrow 4$ $A_2(0)$ CH$_4$ transition at 3057.687423 cm$^{-1}$. CH$_4$ is introduced to the intracavity cell at flow rates from 0.01 to 15 sccm, along with 100 sccm helium buffer gas. As the CH$_4$ flow is increased, the resonant cavity fringe is observed to broaden, then split (Figure 3a). This peak splitting, which we ascribe to the Rabi splitting $\Omega_R$, increases with increasing CH$_4$ flow. The linewidths of the split peaks appear to be set by the cavity mode linewidth, as we expect for polariton peaks,[42] rather than by the inhomogeneously-broadened molecular linewidth. We plot this behavior over a wider range of CH$_4$ flow rates in Figure S3. In Figure S4, we show that neighboring off-resonance cavity fringes remain largely unperturbed by the addition of intracavity CH$_4$. Figure S5 shows that the observed mode splitting is independent of the incident laser power used to monitor the cavity transmission.

We simulate our cavity transmission spectra in the presence of a strong intracavity absorber using the classical expression for the intensity of light transmitted through a Fabry-Pérot cavity with two identical mirrors, given by:[43,44]

$$\frac{I_T(\nu)}{I_0} = \frac{T^2 e^{-\alpha(\nu)L}}{1+R^2 e^{-2\alpha(\nu)L}-2R e^{-\alpha(\nu)L}\cos\left(\frac{4\pi L n(\nu)\nu}{c}\right)} \quad (1)$$

where $L$ is the cavity length and $R$ and $T$ are the reflectivity and transmission of a single mirror. We obtain the frequency-dependent absorption coefficient, $\alpha(\nu)$, and refractive index, $n(\nu)$, of CH$_4$ using data from HITRAN[45] processed with the PGOPHER software package.[46] We perform a least-squares fit of each experimental trace in Figure 3a using eq 1 to arrive at the simulated cavity transmission spectra shown in Figure 3b, reaching excellent agreement. This fitting procedure is described further in Section S2 of the SI. Our fits permit extraction of the CH$_4$ number density corresponding to each flow rate. From the simulated spectra, we additionally extract the mode splitting at resonance, which we plot as a function of [CH$_4$] in Figure 3c. We highlight the [CH$_4$]$^{1/2}$ functional dependence of this splitting in Figure 3c, which is characteristic of the collective strong light-matter coupling regime.[9] We observe a maximum mode splitting of 454 MHz at a CH$_4$ flow rate of 15 sccm, or [CH$_4$] = 3.5 × 10$^{15}$ cm$^{-3}$. We can pursue larger Rabi splittings by further increasing [CH$_4$], though in practice we expect the CBGC temperature to become unstable at high gas flow rates.

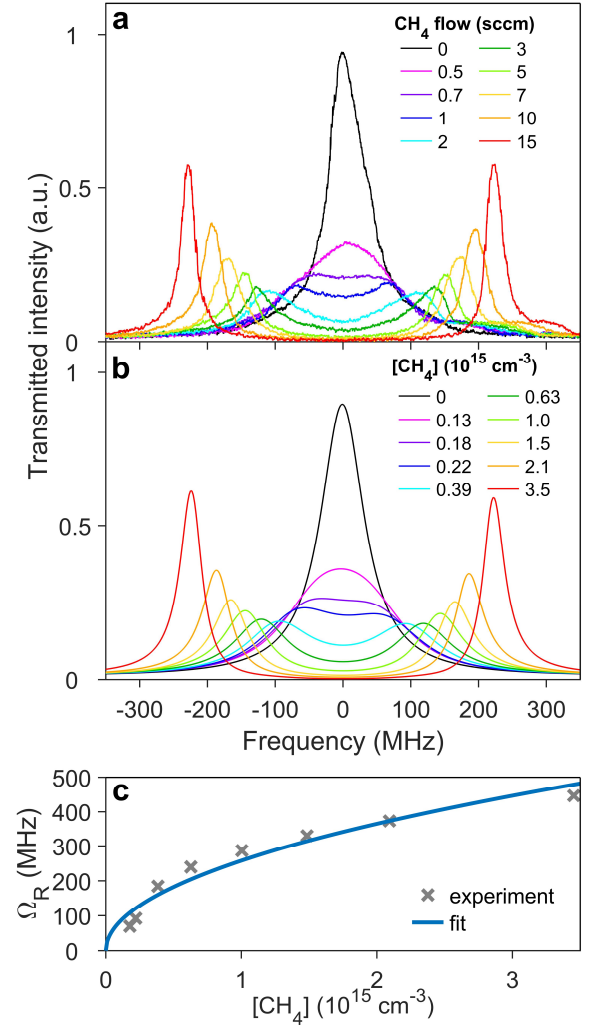

**Figure 3**. (a) Experimental transmission spectra of a near-confocal Fabry-Pérot cavity under systematic tuning of the intracavity CH$_4$ flow rate. In all traces, the cavity mode is kept locked within 10 MHz of resonance with the CH$_4$ $\nu_3$ $J = 3 \rightarrow 4$ $A_2(0)$ transition at 3057.687423 cm$^{-1}$. The frequency axis (300 MHz = 0.01 cm$^{-1}$) is referenced with this resonance corresponding to 0 MHz. The traces are normalized against the amplitude of a fringe far from resonance. (b) Simulated cavity transmission spectra obtained through fitting of the corresponding experimental traces to the classical expression given in eq 1. (c) The Rabi splitting, $\Omega_R$, is extracted from the simulated spectra and plotted as a function of [CH$_4$] (gray), along

with a fitted function (blue) highlighting the $[CH_4]^{1/2}$ functional dependence.

We next confirm that cavity-coupled $CH_4$ behaves as expected under systematic detuning of the cavity mode from resonance with the target transition. Our cavity-locking scheme allows us to stabilize the cavity length at any arbitrary detuning, as shown in Figure S6. We systematically step the cavity length and acquire cavity transmission spectra at each detuning; the resulting experimental dispersion plot is shown in Figure 4a. Individual detuning traces are also shown in Figure S7. We observe the characteristic avoided crossing as the cavity fringe passes through resonance with the target $\nu_3$ $J = 3 \to 4$ $A_2(0)$ transition of $CH_4$. A secondary avoided crossing due to a higher-order mode coupling to the $\nu_3$ $J = 3 \to 4$ $F_2(0)$ transition is also visible in the upper right-hand corner of Figure 4a. In Figure 4b, we use eq 1 to closely reproduce our detuning results. This system therefore behaves consistently with prior demonstrations of collective strong coupling in cold atoms[36,37] and condensed-phase molecules.[23,47]

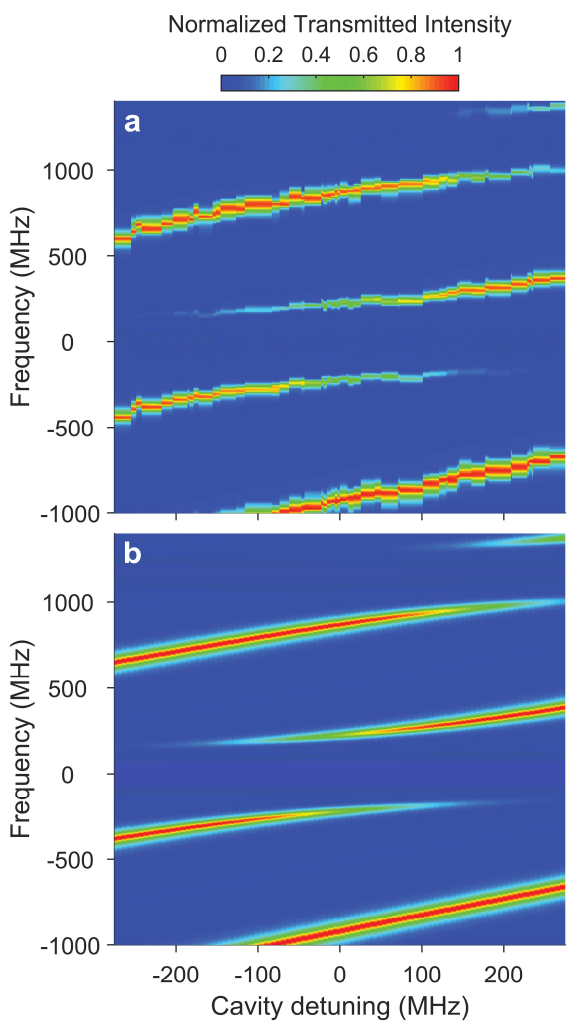

**Figure 4**. (a) Dispersion plot of the experimental transmission spectra of the near-confocal cavity containing a flow rate of 15 sccm of $CH_4$. The cavity is systematically detuned from resonance by varying its length. The frequency axis (300 MHz = 0.01 cm$^{-1}$) is referenced with the $\nu_3$ $J = 3 \to 4$ $A_2(0)$ $CH_4$ transition at 3057.687423 cm$^{-1}$ corresponding to 0 MHz. (b) Simulated dispersion plot of the cavity transmission spectra obtained with eq 1.

In summary, we report the first demonstration of vibrational polaritons in gas-phase molecules. Our apparatus provides unique advantages for future studies of cavity-altered molecular processes, complementary to the planar microfluidic cavity infrastructure typically harnessed for condensed-phase studies. Our centimeter-scale open cavity operates in a distinct experimental regime from wavelength-scale microcavities and allows spectroscopic access from the side to permit direct interrogation of intracavity dynamics. This platform also permits active feedback-stabilization of the cavity length. Whereas microcavities support a continuous spectrum of transverse photonic modes with non-zero in-plane momenta, our near-confocal Fabry-Pérot cavity has a discrete transverse spatial mode structure and conveniently degenerate spatial modes.

This platform may be readily applied to couple rovibrational transitions in other molecular gases or to couple rovibrationally-resolved electronic transitions in future efforts. The extension to room temperature gas-phase VSC should also be feasible. Our cavity design features easily-modifiable length, mirror curvature, and mirror coatings and is also adaptable, for instance, to much higher-finesse or smaller-volume geometries. The single-molecule cavity coupling strength will increase for smaller mode volumes, provided the Rabi splitting is kept fixed. It is possible that stronger coupling per molecule would result in more pronounced polaritonic effects in smaller cavities. A smaller mode volume cavity could also accommodate coupling of species in a molecular beam. We anticipate that this infrastructure will enable exploration of a range of molecular processes under strong coupling in the absence of solvent effects like inhomogeneous local environments and rapid energy dissipation to the bath, and ultimately provide new benchmarks for accurate theoretical treatment of polariton chemistry.

## ASSOCIATED CONTENT

**Supporting Information.** This material is available free of charge via the Internet at http://pubs.acs.org.
Experimental methods, additional experimental cavity transmission spectra and simulations. (PDF)

## AUTHOR INFORMATION

### Corresponding Author

* Marissa L. Weichman - Department of Chemistry, Princeton University, Princeton, New Jersey 08544, United States; https://orcid.org/0000-0002-2551-9146;  Email: weichman@princeton.edu

### Authors

Adam D. Wright - Department of Chemistry, Princeton University, Princeton, New Jersey 08544, United States; https://orcid.org/0000-0003-0721-7854
Jane C. Nelson - Department of Chemistry, Princeton University, Princeton, New Jersey 08544, United States; https://orcid.org/0000-0002-2560-2775

### Author Contributions

The manuscript was written through contributions of all authors. All authors have given approval to the final version of the manuscript.

### Funding Sources

M. L. W. acknowledges support from the ACS Petroleum Research Fund under project PRF-62543-DNI6 and startup funds provided by Princeton University.

### Notes

The authors declare no competing financial interests

## ACKNOWLEDGMENTS

The authors acknowledge the use of Princeton's Imaging and Analysis Center (IAC), which is partially supported by the Princeton Center for Complex Materials (PCCM), a National Science Foundation (NSF) Materials Research Science and Engineering Center (MRSEC; DMR-2011750). This research made use of the PRISM Cleanroom at Princeton University. The authors thank Victoria M. Zhang for assistance with implementing the cavity stabilization scheme.

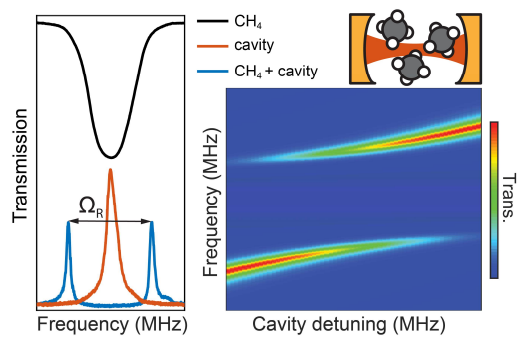

# Supporting Information:
# Rovibrational Polaritons in Gas-Phase Methane

*Adam D. Wright, Jane C. Nelson, Marissa L. Weichman\**

Department of Chemistry, Princeton University, Princeton, New Jersey 08544, United States

\* weichman@princeton.edu

# Contents

# S1. Experimental methods

## S1.1. Cryogenic buffer gas cell

Our home-built cryogenic buffer gas cell (CBGC) and vacuum chamber are shown in Fig. S1. We evacuate the chamber with a turbo pump (Pfeiffer) backed by a roughing scroll pump (Edwards). The cell is mounted to the bottom of a liquid nitrogen (LN$_2$) dewar with ~1.5 L capacity. An LN$_2$ dewar is experimentally simpler than a helium cryostat for this initial demonstration and provides much greater cooling power, enabling thermalization of high number densities of warm molecules.

The dewar provides some cryopumping when filled with LN$_2$, and the chamber reaches a base pressure of $1.1\times10^{-5}$ torr when cold. Room temperature methane gas (Airgas) is delivered to the front aperture of the CBGC. The front aperture is surrounded by an annular slit inlet,[1,2] through which helium buffer gas (Airgas) is introduced. The helium is pre-cooled to the cell wall temperature and thermalizes the methane ensemble through collisions. The flow of both gases is controlled by mass flow controllers (Alicat Scientific). We use a helium flow of 100 sccm for all measurements, while the CH$_4$ flow is varied between 0 and 15 sccm. Optical access to the chamber is made possible through CaF$_2$ flats mounted at the Brewster angle, in line with the in-vacuum optical cavity access described below.

We monitor the dewar and cell temperatures with K-type thermocouples (Omega). Both thermocouples read 90 K before gas flow, and up to 110 K during gas flow. We report a 120 K translational temperature of the methane based on the observed Doppler linewidths (see main text), which is slightly warmer than the cell wall temperature. We ascribe this imperfect thermalization to the fact that methane is volatile, and its freezing and condensation points (91 K and 112 K respectively) are close to the cell wall temperature during our experiments. This is in contrast to the typical use case of CBGCs to study low-vapor pressure molecules that stick to the cell walls and are lost from measurement upon collision.[1-3]

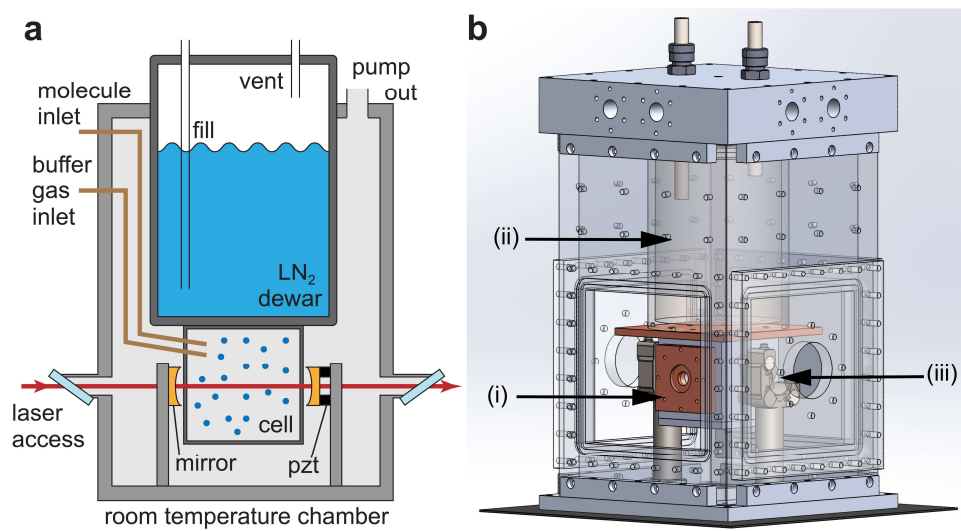

**Figure S1.** Cryogenic buffer gas cell used to prepare a cold, dense molecular sample shown in **(a)** cross-section and **(b)** as a 3D CAD model. The cryogenic cell (i) is cooled by a liquid nitrogen (LN$_2$) dewar (ii) and enclosed by the cavity optics (iii).

## S1.2. Optical cavities

We construct in-vacuum, two-mirror Fabry-Pérot optical cavities by mounting mirrors on either side of the CBGC in standard optical mirror mounts (Thorlabs Polaris) clamped to the floor of the chamber 8.36 cm apart (Fig. S1b). We fabricate our mirrors in the Princeton Cleanroom using an electron beam evaporator (Angstrom Engineering Nexdep) to deposit gold on the concave faces of plano-concave 1" diameter $CaF_2$ substrates. We use −8.36 cm radius of curvature (ROC) substrates (EKSMA Optics) for the near-confocal cavity described in the main text. For the mirror coating, we use a nominal thickness of 11 nm of gold to achieve the desired reflectivity. This reflectivity was chosen to yield cavity linewidths comparable in breadth to those of the molecular features to be coupled. Although matching the cavity and molecular lineshapes ensures cavity coupling of all velocity classes of molecules in the ensemble,[4] here we aim for cavity linewidth slightly narrower than the molecular transitions in question to improve the spectral resolution of the Rabi splitting. We measure a mirror reflectivity of 87.8% at 3270.4 nm (3057.7 cm$^{-1}$) using an infrared microscope (Thermo Scientific Nicolet iN10 MX). These mirrors yield a cavity with a finesse of ~24, consistent with the observed cavity linewidths described in the main text.

For the optical cavity used here, with length $L$ = 8.36 cm and mirrors with ROC = −8.36 cm, the Gaussian $TEM_{00}$ mode will have a $1/e^2$ waist diameter of 0.42 mm and a corresponding mode volume of approximately $7.7 \times 10^{-3}$ cm$^3$. At the highest methane number density that we report here, $[CH_4] = 3.5 \times 10^{15}$ cm$^{-3}$, the number of methane molecules residing in this mode volume will therefore be $2.7 \times 10^{13}$. At a temperature of 120 K, 10.8% of these molecules will occupy the correct $J = 3$ rovibrational level to resonantly couple to the cavity. We therefore estimate that ~$2.9 \times 10^{12}$ methane molecules are available to couple to the lowest-order spatial mode of a given near-resonant longitudinal mode.

## S1.3. Mid-infrared spectroscopy

Our laser table layout is shown in Fig. S2. We use a continuous-wave distributed feedback interband cascade laser (ICL, Nanoplus) to measure the mid-infrared methane and optical cavity transmission spectra. The ICL is designed to operate near the 3270.4 nm (3057.7 cm$^{-1}$) target transition of $CH_4$. Its central wavelength can be tuned over 3262 to 3278 nm (3065 to 3050 cm$^{-1}$). It features a sub-10 MHz instantaneous linewidth[5] and several milliwatts of output power. We control the ICL output wavelength using a current driver (Newport) and a temperature controller (Thorlabs). The ICL is protected from optical feedback by an optical isolator (Faraday Photonics). The ICL features a frequency modulation input port, allowing us to sweep its frequency using a triangle wave from a function generator (BK Precision) to collect spectra.

The ICL light is split into three beams using mid-IR beam splitters (Thorlabs). One beam (labeled 1 in Fig. S2) is aligned through the CBGC chamber to measure the cavity transmission, while the other two beams are used for absolute and relative frequency calibration (see Section S1.4. below). For the CBGC beam path, we use plano-convex $CaF_2$ lenses to mode-match the beam to the $TEM_{00}$ cavity mode. We measure the intensity of light transmitted through the cavity as a function of sweep frequency with an MCT detector (Kolmar). We measure the absorption spectrum of methane in the CBGC by removing the cavity mirrors.

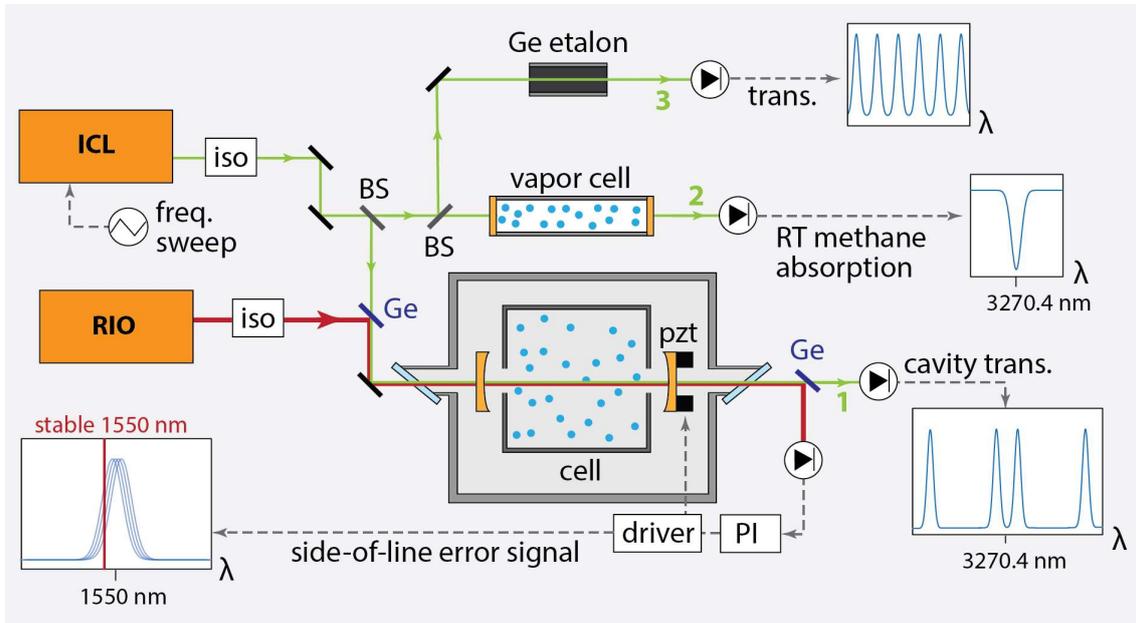

**Figure S2.** Laser scheme used to record the cavity and cryogenic cell transmission spectra, calibrate the interband cascade laser (ICL) frequency, and stabilize the optical cavity length. The 3270.4 nm ICL beam (red) is divided into three parts by beam splitters (BS). The cavity transmission (1) and calibration spectra (2,3) are measured by sweeping the ICL frequency. We use a metrology-grade diode laser (RIO) to actively stabilize the cavity length via a side-of-line lock to a cavity fringe at a wavelength near 1550 nm. The locking error signal is processed with a proportional-integral (PI) loop filter and fed back onto a piezoelectric chip (pzt). Abbreviations: iso = isolator, RT = room temperature.

## S1.4. Laser frequency calibration

We calibrate the frequency of the ICL laser output using a homebuilt vapor cell containing room-temperature methane as an absolute frequency reference (beam 2 in Fig. S2), and a germanium (Ge) etalon (Light Machinery, nominal length 50.8 mm) as a relative reference (beam 3 in Fig. S2). The transmission of the ICL light through both vapor cell and etalon is captured with InAsSb detectors (Thorlabs), and recorded synchronously with the CBGC cavity transmission data as the ICL frequency is swept.

The three $CH_4$ absorption features within the range of the laser sweep (see Fig. 2b) provide absolute frequency reference points, as their peak positions are known with high precision.[6] In order to establish the laser frequency relative to these features, we perform fringe-counting using the Ge etalon transmission spectrum. We extract the temporal positions of the extrema of the etalon transmission spectrum as the laser frequency is scanned and fit them with a second-order polynomial. This polynomial accounts for nonlinearities in the ICL frequency response to the triangle wave current sweep. We use this polynomial to rescale the frequency sweep axis such that the etalon extrema are evenly spaced by the appropriate free spectral range (FSR). We determine the etalon FSR as 719.5 MHz, which provides the best agreement of the rescaled frequency axis with the $CH_4$ reference features. This method calibrates our frequency axis with sufficient accuracy that spacings between the methane absorption peaks within the laser sweep range have <10 MHz error.

## S1.5. Cavity stabilization

We have implemented a cavity length stabilization scheme to allow us to lock a cavity fringe on-resonance or tunably off-resonance with the desired molecular transition. One in-vacuum cavity mirror is glued to a piezoelectric ring chip (Thorlabs) to actuate the cavity length. We stabilize the cavity length using a side-of-line lock with a cavity fringe as a frequency discriminator (see Fig. S2). As the cavity fringes near the 3270.4 nm ICL wavelength shift in frequency in the presence of strongly-absorbing intracavity methane, we perform the side-of-line lock in a separate near-IR wavelength regime, far from resonance with any methane lines. We use a metrology-grade continuous wave laser (RIO ORION) operating near 1550 nm as our frequency reference. The RIO output is aligned through the optical cavity, concentric with the ICL beam. The two beams are overlapped before the cavity and separated after cavity transmission using Ge windows with mid-IR antireflection coating (Thorlabs) as dichroic mirrors, which transmit the ICL light while reflecting the 1550 nm beam with good discrimination. The intensity of 1550 nm light transmitted through the cavity is detected with an amplified Ge detector (Thorlabs), which serves as the side-of-line error signal. We process this error signal using a proportional-integral loop filter (New Focus) and feed it back onto the cavity piezo using a current driver (Thorlabs). To systematically step the cavity length for detuning measurements, we step the output wavelength of the RIO laser via the thermistor resistance of its thermoelectric cooler. The locked cavity then follows the laser to stabilize at a new cavity length. This scheme ensures that each cavity transmission measurement is robust against vibrational noise and long-term drift.

# S2. Simulation of cavity transmission spectra

We simulate cavity transmission spectra using the classical optics expression for the fraction of light transmitted through a two-mirror Fabry-Pérot cavity.[7,8] We present this expression in the main text (eq 1), and describe it again here in more detail.

$$\frac{I_T(v)}{I_0} = \frac{T^2 e^{-\alpha(v)L}}{1 + R^2 e^{-2\alpha(v)L} - 2R e^{-\alpha(v)L}\cos(\frac{4\pi L n(v) v}{c})} + \text{p.s.} \quad \text{(S1)}$$

where $L$ is the cavity length, $R$ and $T$ are the reflectivity and transmission coefficients for a single cavity mirror, $\alpha(v)$ and $n(v)$ are the absorption coefficient and refractive index of the intracavity medium, $c$ is the speed of light, and $v$ is the laser frequency. Unlike other cavity geometries, a near-confocal cavity features degenerate transverse modes, and features a longitudinal mode spacing that is half that of the cavity free spectral range.[7] To capture this denser mode spacing for the near-confocal case, we include a "phase shifted" term in eq S1 designated p.s., which is identical to the first term except with the addition of a phase shift of $\pi$ within the argument of the cosine function.

To simulate our experiments, we calculate $\alpha(v)$ and $n(v)$ for $CH_4$ at 120 K from the absorption cross section $\sigma(v)$, which is obtained from HITRAN data[6] processed with the PGOPHER program.[9] We obtain the absorption coefficient from the absorption cross section according to the Beer-Lambert law:

$$\alpha(v) = \frac{N}{V}\sigma(v) \quad \text{(S2)}$$

where $N/V$ is the $CH_4$ number density. The imaginary component of the refractive index, or extinction coefficient, $\kappa(v)$, can be obtained from the absorption coefficient according to:[10]

$$\kappa(v) = \alpha(v)\frac{c}{4\pi v} \quad \text{(S3)}$$

Finally, we recover the real component of the refractive index, $n(v)$, from $\kappa(v)$ via the Kramers-Kronig relation.

We use eq S1 to fit the experimental cavity transmission spectra via a least-squares minimization method, with $L$, $N/V$, $R$, and $T$ as fitting parameters. For the simulated spectra shown in Fig. 3b of the main text, $R$ was fit globally across the spectra taken with different $CH_4$ number densities. A value of $R=0.87$ was found to be the best fit, consistent with our experimentally measured mirror reflectivities (see Section S1.2 above). Once $N/V$ was obtained from the fit for each spectrum, $L$ was varied to simulate cavity transmission spectra under perfect resonance conditions, when the split peaks were symmetrical in intensity. The Rabi splittings, $\Omega_R$, plotted in Fig. 3c of the main text were calculated from the separation of the peak maxima of the on-resonance simulated spectra. When the peak maxima were not distinct, this splitting was inferred from the separation of the maxima of the third derivative of the transmitted intensity.

# S3. Additional figures

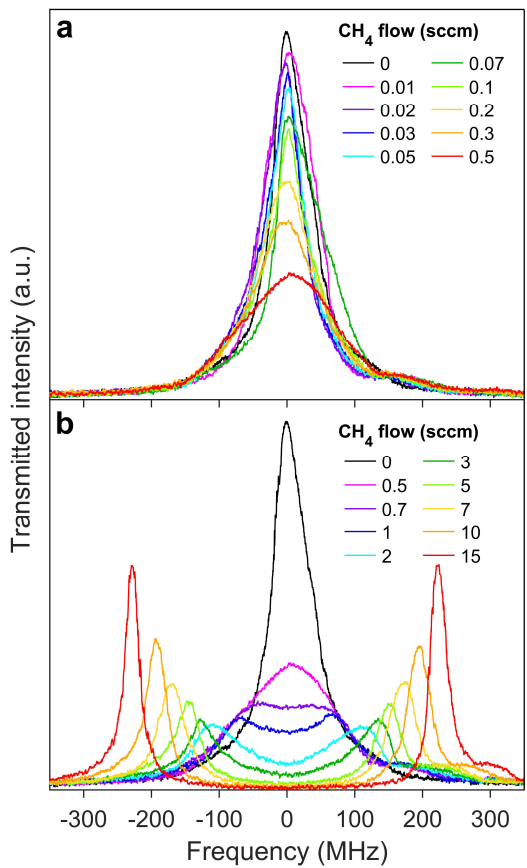

**Figure S3.** Experimental transmission spectra of the near-confocal Fabry-Pérot cavity under systematic tuning of the intracavity $CH_4$ flow rate for **(a)** lower and **(b)** higher $CH_4$ flow rates. In all traces, the cavity mode is kept locked within 10 MHz of resonance with the $CH_4$ $\nu_3$ $J = 3 \rightarrow 4$ $A_2(0)$ transition at 3057.687423 cm$^{-1}$. The frequency axis (300 MHz = 0.01 cm$^{-1}$) is referenced with this resonance corresponding to 0 MHz. As the $CH_4$ flow rate is increased, the cavity transmission feature first decreases in amplitude, before exhibiting Rabi splitting for flow rates larger than ~1 sccm.

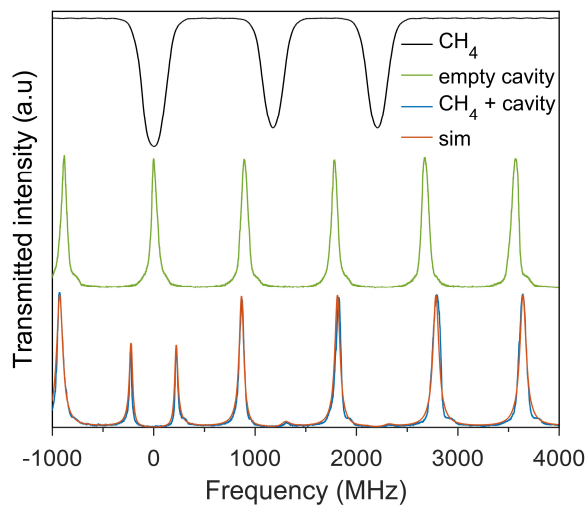

**Figure S4.** Transmission spectrum of the evacuated near-confocal cavity (green) and of the same cavity under an intracavity 15 sccm flow of $CH_4$ (blue), with a cavity mode locked on resonance with the main $CH_4$ transition of interest at 3057.687423 cm$^{-1}$. The simulated transmission spectrum (orange) closely reproduces the experimental system over a wide frequency window. The 120 K $CH_4$ transmission spectrum is plotted in black and vertically offset for clarity. We observe minor shifting of neighboring cavity modes caused by the additional $CH_4$ transitions at 3057.726496 and 3057.760735 cm$^{-1}$.

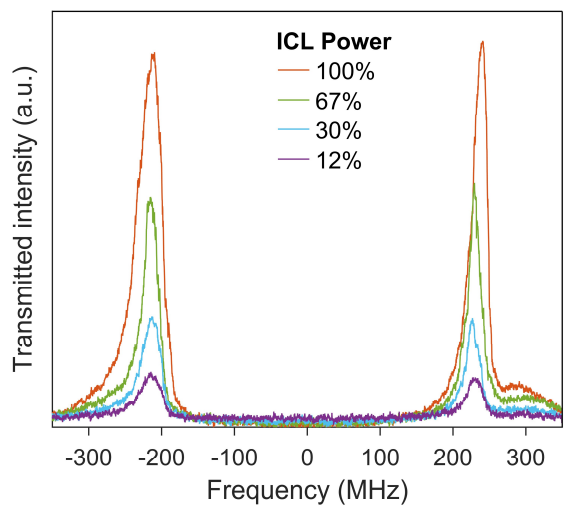

**Figure S5.** Transmission spectra of the near-confocal cavity under a 15 sccm flow of $CH_4$ and with a range of incident mid-IR laser powers. 100% laser power corresponds to 430 μW incident on the input window of the vacuum chamber. As we decrease the laser power, the transmitted signal correspondingly drops in intensity, but the peak positions remain consistent. For each trace, the cavity fringe was locked on resonance with the $CH_4$ transition at 3057.687423 cm$^{-1}$. The frequency axis is referenced with this resonance corresponding to 0 MHz.

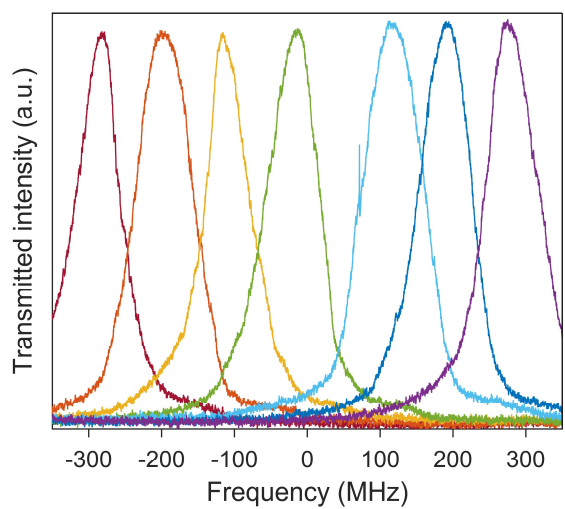

**Figure S6.** Transmission spectra of an evacuated near-confocal cavity over a range of cavity detunings. In each case, the cavity length is stabilized with the side-of-line lock described in Section S1.5. The frequency axis is referenced with the $CH_4$ transition at 3057.687423 cm$^{-1}$ corresponding to 0 MHz.

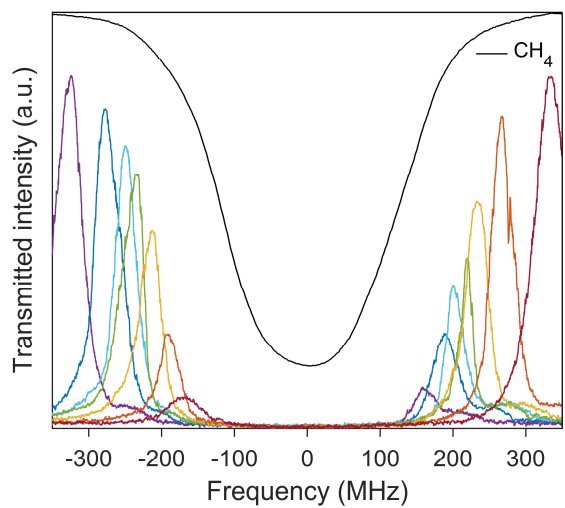

**Figure S7.** Transmission spectra of the near-confocal cavity, under a 15 sccm flow of $CH_4$ and with a range of cavity lengths (colored traces). The 120 K spectrum of $CH_4$ is plotted in black and vertically offset for clarity. As the cavity mode is brought into resonance with the $CH_4$ transition, the cavity mode splitting reaches a minimum and the intensities of the split peaks are maximally symmetric (yellow trace). The frequency axis is referenced with the $CH_4$ transition at 3057.687423 cm$^{-1}$ corresponding to 0 MHz. These traces are used to construct the 2D dispersion plot in Fig. 4 of the main text.